# Continuum versus Discrete: A Physically Interpretable General Rule For Cellular Automata By Means of Modular Arithmetic

Luan Carlos de Sena Monteiro Ozelim

*University of Brasília, Brasília - DF, Brazil, Department of Civil and Environmental Engineering, Zip Code 70910-900, E-mail: luanoz@gmail.com*

André Luís Brasil Cavalcante

*University of Brasília, Brasília - DF, Brazil, Department of Civil and Environmental Engineering, Zip Code 70910-900, E-mail: abrasil@unb.br*

Lucas Parreira de Faria Borges

*University of Brasília, Brasília - DF, Brazil, Department of Civil and Environmental Engineering, Zip Code 70910-900, E-mail: lucaspdfborges@gmail.com*

Abstract: Describing complex phenomena by means of cellular automata (CA) has shown to be a very effective approach in pure and applied sciences. In fact, the number of published papers concerning this topic has tremendously increased over the last twenty years. Most of the applications, notwithstanding, use cellular automata to qualitatively describe the phenomena, which is surely a consequence of the way the automata rules have been defined. In the present paper a general rule which describes every of Wolfram's cellular automata is derived. The new representation is given in terms of a new function hereby defined, the iota-delta function. The latter function is further generalized in order to provide a general rule for not only Wolfram's but also to every CA rule which depends on the sum and products of the values of cells in the automaton mesh. By means of a parallel between the finite difference method and the iota-delta function, the new representation provides a straightforward physical interpretation of CA, which gives, for the first time, a quantitative interpretation of the generating rule itself. By means of the new formulation, advective-diffusive phenomena are analyzed. In particular, the relation between CA automata and anomalous diffusion is briefly discussed.

# Introduction

It is undeniable that Science has evolved to such a stage that almost every situation in mankind every-day life has been, in some way, addressed. This can be clearly seen when, for example, technological development is taken into account. Such development provided people better conditions to improve and build the society as one knows today.

One may not forget, on the other hand, that knowledge and technology are intimately related. The most up-to-date gadgets employ state-of-the-art concepts of Physics, Mathematics and other basic sciences. This way, the need to keep researching in basic sciences is inherent to the development of the so-called applied sciences, such as Engineering.

This rigid segregation of the production and application of knowledge started changing when every-day problems complexity increased. Both basic and applied sciences had to strengthen bonds in order to give satisfactory answers to the appearing problems. In special, the atomic theory played a major role in this need of jointly work.

At a given time, some notable physicists, such as Werner Heisenberg, Niels Bohr, Max Plank, Erwin Schrödinger and Louis de Broglie noticed that, even if they correctly applied the established knowledge, no plausible answer would come out for their questions about atomic structure. It is worth asking whether the questions of some of the greatest minds of our era were misformulated or the main issue was not how questions were formulated, but what kinds of answers were expected. Schrödinger (1966) gave a straightforward explanation to this apparent "anomaly" in science, as show in (Goldfarb, 2010, emphasis added):

"[…] If you envisage the development of physics in the last half-century, you get the impression that the discontinuous aspect of nature has been forced upon us very much against our will. We seemed to feel quite happy with the continuum. Max Planck was seriously frightened by the idea of a discontinuous exchange of energy […] Twenty-five years later the inventors of wave mechanics indulged for some time in the fond hope that they have paved the way of return to a classical continuous description, but again the hope was deceptive. Nature herself seemed to reject continuous description […] The observed facts appear to be repugnant to the classical ideal of continuous description in space and time. […] So the facts of observation are irreconcilable with a continuous description in space and time […]"

Also, Fritjof Capra (Capra, 1975, emphasis added) brilliantly stated in his book:

"[…] Every time the physicists asked nature a question in an atomic experiment, nature answered with a paradox, and the more they tried to clarify the situation, the sharper the paradoxes became. It took them a long time to accept the fact that these paradoxes belong to the intrinsic structure of atomic physics, and to realize that they arise whenever one attempts to describe atomic events in the traditional terms of physics. […]"

Both the excerpts above can be readily exemplified by one of the most remarkable and, in some ways, unintentional changes in scientific ideas: the quantization of energy by Max Plank. While studying the second law of thermodynamics, both Plank and Ludwig Boltzmann had a serious rivalry. The former, at first, did not believe that entropy would be statistically described, while the latter firmly defended such interpretation (Kragh, 2000).

In order to give a consistent explanation to the increase of entropy predicted by thermodynamics, Plank and his contemporaries deeply analyzed Maxwell's laws since these relations were supposed to govern the microscopic oscillators that produced the heat radiation emitted by black bodies (Kragh, 2000). After some time, Plank believed he had justified the irreversibility, thus the entropy change, by means of the lack of symmetry of Maxwell's equations. Boltzmann promptly showed that Plank was wrong, impelling the latter to seek another way to explain the second law of thermodynamics (Kragh, 2000).

While studying the black body radiation emission problem, Plank came up with a theoretically justifiable formula which matched well experimental results. Sir Rayleigh and Sir James Jeans, at about the same time, proposed an energy distribution based on classical mechanics, i.e., continuum theory. The so-called Rayleigh-Jeans law (R-JL) led to a classical physical misinterpretation: the ultraviolet catastrophe. Following R-JL, the production of energy was proportional to the inverse of the fourth-power of the wave length, this way, when ultraviolet radiation is considered, as the wave length decreases, the energy production tends to infinity, which is absurd (Kragh, 2000).

In the end of the year 1900, Plank noticed that the equation he proposed was more than a lucky shot and, in order to give it solid theoretical basis, he had to adopt some of Boltzmann's probabilistic ideas. By doing that, for the first time the so-called Boltzmann equation appeared. In short, the latter equation relates entropy to molecular disorder. In order to quantify molecular disorder, Plank had to establish a way to count the number of ways a given energy can be distributed among a set of oscillators. The creation of the concept of quanta, discrete elements with finite portion of energy, was the answer to his questions (Kragh, 2000).

This way, the determinant issue while asking effective questions to nature is the dichotomy: continuum vs. discrete. The continuity ideas date back to ancient times and seem to have found in Parmenides and Aristotle (Wall, 2011) its first defenders. By Modern Age, Isaac Newton and Gottfried Leibniz established one of the pillars of the continuity principle: differential and integral calculus. Everything that is currently described by means of differential equations has the continuity principle as background. Such relation follows from the intrinsic linkage between differentiability and continuity of functions, leading to the "continuity" of the phenomena described. Scientists got so amazed by differential calculus that Bernhard Riemann once said (Goldfarb, 2010, emphasis added):

"[…] As is well known, physics became a science only after the invention of differential calculus. It was after realizing that natural phenomena are continuous that attempts to construct abstract models were successful. […] True basic laws can only hold in the small and must be formulated as partial differential equations. Their integration provides the laws for extended parts of time and space. […]"

By now, a very important remark has to be taken into consideration: continuity, in fact, provided mankind an outstanding development. As stated in the very first paragraph of the present paper, one could not have come this far without continuity and its models. The purpose of this paper is not to depreciate the continuum ideas, but to propose a better approximation of natures' behavior since nature itself, in Schrödinger's words, seems "to reject continuous description".

Riemann, maybe dazzled by the tremendous success arising from the application of differential equations in science, made a few statements which were, in some ways, trifling. As noticed by Goldfarb (2010), the assumption that natural phenomena were continuous seems to be postulated rather than noticed. An important question is whether the integration referred by Riemann always apply. These questions have found their answers in current scientific community: the growing usage of numerical methods in which finite elements or differences are taken into account show that, once again, one has to treat in a discrete way what was once thought to be continuous.

The basic idea behind finite element method is mesh discretization of some continuous domain into a set of discrete sub-domains. Finite differences method, on the other hand, transforms differential equations into difference equations, being the latter a discrete "approximation" of the former.

If nature shows to behave discretely, and the way found to solve the governing equations of the established continuum theory is by discrete differences and finite elements, it seems the scientific community keeps doubling its work.

It is remarkable how continuous models fit to circular, rectangular, square and, in general, well-defined geometries. But, as the visionary Benoit Mandelbrot said (Mandelbrot, 1982):

"[…] Clouds are not spheres, mountains are not cones, coastlines are not circles, and bark is not smooth, nor does lightning travel in a straight line. […]"

Put in another way, continuous works well when idealized problems are considered, but only discrete methods can bring the answers to complex problems. By complex, on the other hand, one does not mean difficult or unsolvable. Complexity is inherent to nature. Stephen Wolfram, in his paradigm shifting book "A New Kind of Science", clearly showed that simple rules – also referred by Wolfram as programs – can generate complexity and intricate patterns (Wolfram, 2002).

In short, at first, by means of simple binary language, Wolfram proposed a class of computational experiments over a net of cells. The values of three cells determine how a fourth cell would be. Since each cell can have 2 possible states, there are 8 possible trios. Each trio can result in other two values for the fourth cell, thus, there are $2^8$ possibilities. Taking advantage of the binary scenario, Wolfram created the 0-255 classification, in which each combination receives a number from 0 to 255. This will be further explained in the following sections.

The generalization from binary to ternary was immediate and Wolfram also presented it his book. A concernment shown by Wolfram was whether a general simple rule which would govern time, space and energy in a given scale existed or not. In order to address this issue, the present paper shows that Wolfram's cellular automata can be expressed by a single rule applied to the whole cellular net. It is then conjectured that every cellular automata can be represented by a generalization of the referred rule. Also, by means of a parallel between the new general rule hereby defined and the finite difference method, it is possible, for the first time, to quantitatively describe cellular automata and the phenomena by the latter modeled.

## Investigating the 0 – 255 Wolfram's Cellular Automata

In the second chapter of the book "A New Kind of Science" by Wolfram (2002), a simple yet tricky question is asked: "How do Simple Programs Behave?". For sure, answering such question is extremely difficult if one tries to do so merely based on standard science. Prior to being able to predict is to feel, observe, interact. Computers provided mankind the possibility to take contact with multiple realities, allowing the observation of the interaction between multiple phenomena at the same time.

This way, Wolfram (2002) shows that using computers to answer the question referred in the last paragraph is not just a good way to address the issue, but the only method which can bring reliable results. Experimental computation led to the establishment of cellular automata, one of the basis of the new kind of science.

Let one start with the simplest cellular automata available in Wolfram's groundbreaking book (Wolfram, 2002). In short, the classical definition is (Wolfram, 2002):

"The cellular automaton consists of a line of cells, each colored black or white. At every step there is then a definite rule that determines the colour of a given cell from the colour of that cell and its immediate left and right neighbors on the step before."

The excerpt above can be better visualized by considering Figure 1:

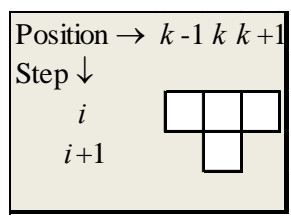

Figure 1. Rule Application Scheme.

As stated by Wolfram, one has to take into account three cells from the last step in order to define the color (or, as it is commonly described, the binary value 1 or 0) of a cell in the next step. In other words, the value of a given cell is given as a function of the values of three other cells. In order to give a mathematical description of this relation, let us consider the following representation:

$$C_k^{i+1} = f(C_{k-1}^i, C_k^i, C_{k+1}^i) \qquad (1)$$

in which $C_k^i$ is the value assumed by the cell in position *k* at step *i*. Since there are only two possible values for the cells, either 1 – black – or 0 – white –, there are $2^3$ possible combinations or trios, as seen in Figure 2:

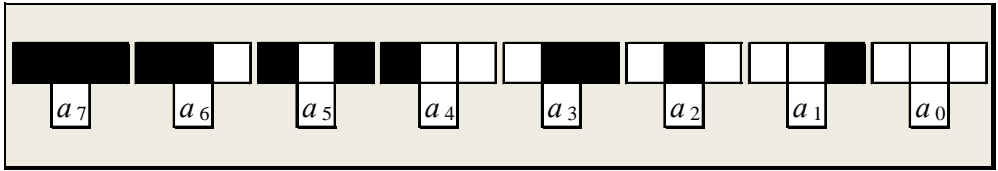

Figure 2. The Eight Possible Trios.

Each value $a_j$, $j = 0, 1,...,7$, is equivalent to 1 or 0, this way, a total of $2^8$ possible rules are determined by the simple procedure described above. Wolfram established a naming criteria for all this 256 possible trio combinations known as the 0 – 255 classification (Wolfram, 2002). The rule is straightforward and relates the decomposition of the rule number in base 2 to the values of the coefficients $a_j$. For example, the rule 30 has the following factorization:

$$30 = 0.2^7 + 0.2^6 + 0.2^5 + 1.2^4 + 1.2^3 + 1.2^2 + 1.2^1 + 0.2^0 \qquad (2)$$

The coefficients $a_j$, are the numbers that multiply $2^j$ in the factorization process. This way, the set of trios which generate rule 30 is described in Figure 3:

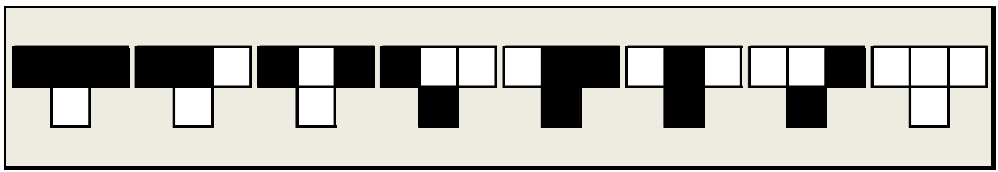

Figure 3. Rule 30 Trios Combination.

In order to produce a compact notation, let one define the Rule Number function (RNf) as:

$$RNf(a_0, a_1, a_2, a_3, a_4, a_5, a_6, a_7) = \sum_{j=0}^{7} a_j 2^j \qquad (3)$$

The RNf provides the rule number given the values of the coefficients $a_j$, $j = 0, 1,...,7$. This way, the logic is inverse: the function inputs are the coefficients and not the rule number. The latter is, on the other hand, the output of the transformation. The RNf can be generalized to a base *b* and a combination of *n* cells is the previous step as:

$$RNf\left(v_0, v_1, v_2, ..., v_{b^n-1}; b\right) = \sum_{j=0}^{b^n-1} v_j b^j; \quad 0 \leq RNf \leq b^{b^n} - 1 \tag{4}$$

The proof is suppressed due to the simplicity inherent to the obtention of Eq.(4).

Given this brief introduction about cellular automata, let us consider another interesting rule: rule 90. Its trio representation is given in Figure 4:

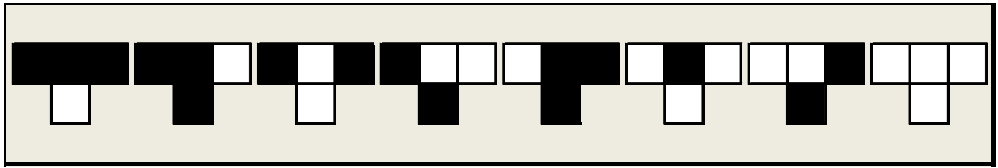

Figure 4. Rule 90 Trios Combination.

In (Wolfram, 2002, adapted) the following relation was given to describe rule 90's behavior:

$$C_k^{i+1} = \text{mod}\left[C_{k-1}^i + C_{k+1}^i; 2\right] \tag{5}$$

in which mod[*o*; *p*] denotes de modulo operator, which gives the rest of the division of *o* by *p* if *o* is greater than *p* or *o* itself, otherwise.

An interesting characteristic of Eq.(5) is its simplicity. Instead of being necessary 8 different rules, as shown in Fig. 4, only one rule can be applied to the whole net in order to get the desired pattern. Using the notation introduced in Eq.(1), Fig.4 gives:

$$C_k^{i+1} = \begin{cases} 0, & \text{if } (C_{k-1}^i, C_k^i, C_{k+1}^i) = (1,1,1) \\ 1, & \text{if } (C_{k-1}^i, C_k^i, C_{k+1}^i) = (1,1,0) \\ 0, & \text{if } (C_{k-1}^i, C_k^i, C_{k+1}^i) = (1,0,1) \\ 1, & \text{if } (C_{k-1}^i, C_k^i, C_{k+1}^i) = (1,0,0) \\ 1, & \text{if } (C_{k-1}^i, C_k^i, C_{k+1}^i) = (0,1,1) \\ 0, & \text{if } (C_{k-1}^i, C_k^i, C_{k+1}^i) = (0,1,0) \\ 1, & \text{if } (C_{k-1}^i, C_k^i, C_{k+1}^i) = (0,0,1) \\ 0, & \text{if } (C_{k-1}^i, C_k^i, C_{k+1}^i) = (0,0,0) \end{cases} \tag{6}$$

In order to further investigate the reason Eq.(5) is much simpler than Eq.(6), the next section brings up a few properties and definitions about modular arithmetic.

## Discrete Mathematics and Modular Arithmetic

While considering the divisibility of integers, Gauss introduced a simple notation which, in many cases, provides a straightforward manner of solving problems. By means of such notation, he created the theory of congruences, which ended up becoming a new branch of number theory (Apostol, 1976).

Modular congruence can be defined as follows (Apostol, 1976): given integers $w$, $g$ and $s$ with $s > 0$. One says that $w$ is congruent to $g$ modulo $s$, and write:

$$w \equiv g \pmod{s} \tag{7}$$

if $s$ divides the difference $w - g$. The number $s$ is called the modulus of the congruence. A few important properties of modular congruence are (Apostol, 1976): If $w \equiv g \pmod{s}$ and $\omega \equiv \gamma \pmod{s}$, then:

$$wx + \omega y \equiv gx + \gamma y \pmod{s} \tag{8a}$$

$$w\omega \equiv g\gamma \pmod{s} \tag{8b}$$

$$w^n \equiv g^n \pmod{s} \tag{8c}$$

for all integers $x$ and $y$ and positive integers $n$.

The modulo operation, as described in Eq.(5), can be defined as (Boute, 1992):

$$\mod[w; s] = w - \left\lfloor \frac{w}{s} \right\rfloor s \tag{9}$$

in which $\lfloor x \rfloor$ stands for the floor function, which provides the largest integer not greater than $x$. It is worth noticing that Eq.(9) is valid for all real numbers. Besides, congruence modulo can, in some ways, be generalized to real numbers.

# A Transformation Which Relates Cellular Automata to Modular Arithmetic: The Iota-Delta Function

While observing Eq.(5), an immediate general rule which would give the representation of other cellular automata in terms of modulo operation would be:

$$C_k^{i+1} = \mod\left[\alpha_1 C_{k-1}^i + \alpha_2 C_k^i + \alpha_3 C_{k+1}^i; 2\right] \qquad (10)$$

in which $\alpha_j$ are integer coefficients such that $0 \leq \alpha_j \leq 1, j = 1, 2, 3$.

One has to take this latter restriction into account because when $\alpha_j < 0$ or $\alpha_j > 1$, consider that $\alpha_j = 2p + r$, $p \in \mathbb{Z}^*, r \in \mathbb{Z}_+$ and $r \leq 1$ this way, from Eq.(10), taking, for example, $j = 1$, the following holds true:

$$(2p + r)C_{k-1}^i + \alpha_2 C_k^i + \alpha_3 C_{k+1}^i \equiv C_k^{i+1} \pmod{2} \qquad (11)$$

It is also valid that:

$$2p \equiv 0 \pmod{2} \qquad (12)$$

From Eqs. (8), (11) and (12) one shall get:

$$rC_{k-1}^i + \alpha_2 C_k^i + \alpha_3 C_{k+1}^i \equiv C_k^{i+1} \pmod{2} \qquad (13)$$

which proves that the non-redundant values of $\alpha_j$ are obtained when $\alpha_j = \{r \mid r \leq 1, r \in \mathbb{Z}_+\}$.

The rules generated by the combination established in Eq.(10) are summarized in Table 1. By inspection, not only rule 90 but also other 7 cases are simply defined by means of a single rule applied to the whole net.

It is interesting to notice that the formula in Eq.(10) will never describe odd rules since in Wolfram's numeration system, the coefficient which regulates the parity is $a_0$ and the latter depends on three cells whose values are zero. This way, when applying Eq.(10) to any linear combination of three zeroes, the null value is obtained, which implies only even rule numbers. In order to overcome this issue, a fourth coefficient $\alpha_4$ needs to be inserted inside the modulo operator in Eq.(10) generating:

$$C_k^{i+1} = \mod\left[\alpha_1 C_{k-1}^i + \alpha_2 C_k^i + \alpha_3 C_{k+1}^i + \alpha_4; 2\right] \qquad (14)$$

Table 1. Rules Described by Eq.(10).

| Coefficients | | | Rule Number |
|---|---|---|---|
| $\alpha_1$ | $\alpha_2$ | $\alpha_3$ | |
| 0 | 0 | 0 | 0 |
| 1 | 1 | 0 | 60 |
| 1 | 0 | 1 | 90 |
| 0 | 1 | 1 | 102 |
| 1 | 1 | 1 | 150 |
| 0 | 0 | 1 | 170 |
| 0 | 1 | 0 | 204 |
| 1 | 0 | 0 | 240 |

For the same reasons described previously, the values of the coefficient $\alpha_4$ are also either 0 or 1. This way, the number of rules described by the application of Eq. (14) is 16, the double of the ones obtained by applying Eq. (10). Table 2 summarizes the rule numbers obtained by the addition of the fourth coefficient.

Table 2. Rules Described by Eq.(14).

| Coeficients | | | | Rule Number |
|---|---|---|---|---|
| $\alpha_1$ | $\alpha_2$ | $\alpha_3$ | $\alpha_4$ | |
| 0 | 0 | 0 | 0 | 0 |
| 1 | 0 | 0 | 1 | 15 |
| 0 | 1 | 0 | 1 | 51 |
| 1 | 1 | 0 | 0 | 60 |
| 0 | 0 | 1 | 1 | 85 |
| 1 | 0 | 1 | 0 | 90 |
| 0 | 1 | 1 | 0 | 102 |
| 1 | 1 | 1 | 1 | 105 |
| 1 | 1 | 1 | 0 | 150 |
| 0 | 1 | 1 | 1 | 153 |
| 1 | 0 | 1 | 1 | 165 |
| 0 | 0 | 1 | 0 | 170 |
| 1 | 1 | 0 | 1 | 195 |
| 0 | 1 | 0 | 0 | 204 |
| 1 | 0 | 0 | 0 | 240 |
| 0 | 0 | 0 | 1 | 255 |

The application of modular arithmetic fits perfectly the need of simplicity in the description of cellular automata. On the other hand, the number of automata generated by Eq.(14) is 1/16 of the total number of binary automata – also called simple 1D automata by Wolfram (2002).

As it can be seen, the number of automata generated is directly related to how many values the coefficients $\alpha_j$ can assume. Also, while investigating Eq.(10), it has been shown that while considering modulo 2, only the numbers 0 and 1 are non-redundant coefficients. Thus, one has to find a way to generate more combinations of coefficients which are non-redundant.

It can be shown that when modulo $n$ is considered, a total of $n^4$ combinations of four non-redundant coefficients is obtained. Besides, the non-redundant coefficients are described as:

$$\alpha_j = \{r | r \leq n-1, r \in \mathbb{Z}_+\}; j = 1,...,4 \tag{15}$$

This way, when modulo 3 is taken into account; the non-redundant coefficients are 0, 1 and 2, which ultimately generate a total of $3^4$ combinations.

On the other hand, since one is considering binary automata, modulo 3 cannot be applied alone as the outcomes of such operation are not only 0 and 1, but also 2. This way it is not possible to describe binary automata by a rule of the form:

$$C_k^{i+1} = \mod\left[\alpha_1 C_{k-1}^i + \alpha_2 C_k^i + \alpha_3 C_{k+1}^i + \alpha_4; 3\right] \tag{16}$$

Thus, the situation is paradoxically summarized as: in order to describe more cellular automata by a simple rule which uses modulo operator, one has to increase the non-redundant values of the coefficients. This growth can only be obtained by considering modulo operator with respect to greater numbers than 2. Notwithstanding, if one considers modulo operator with respect to integers greater than 2, binary automata cannot be described since the outcomes of the transformation are not only 0 and 1. At this point a very important concept has to be introduced: filtering operators.

In order to preserve the number of non-redundant coefficients obtained by considering modulo operator with respect to greater integers and yet obtain only 0 and 1 as the outputs of the transformation, one has to apply mod [,2] to the right-hand side of Eq.(16). This process is a filtering processes in which the results from Eq.(16) are filtered in order to obtain binary outputs. This way, Eq.(16) becomes:

$$C_k^{i+1} = \text{mod}\left[\text{mod}\left[\alpha_1 C_{k-1}^i + \alpha_2 C_k^i + \alpha_3 C_{k+1}^i + \alpha_4 ; 3\right] ; 2\right] \tag{17}$$

It can be noticed that the number of combinations of the non-redundant coefficients in Eq.(17) are $3^4$ and the outputs of the latter equation are only 0 and 1.

One has to pay close attention to the fact that each automaton is not generated by a single combination. Due to the cyclic property of the modulo operator, more than one combination generates the same automaton. This can be verified in Table 3, which shows all the combinations of non-redundant coefficients in Eq.(17) and the correspondent rules generated.

Table 3. Rules Described by Eq.(17).

| Coefficients | | | | RNf | Coefficients | | | | RNf | Coefficients | | | | RNf |
| --- | --- | --- | --- | --- | --- | --- | --- | --- | --- | --- | --- | --- | --- | --- |
| $\alpha_1$ | $\alpha_2$ | $\alpha_3$ | $\alpha_4$ | | $\alpha_1$ | $\alpha_2$ | $\alpha_3$ | $\alpha_4$ | | $\alpha_1$ | $\alpha_2$ | $\alpha_3$ | $\alpha_4$ | |
| 0 | 0 | 0 | 0 | 0 | 2 | 1 | 2 | 0 | 36 | 1 | 1 | 1 | 1 | 129 |
| 0 | 0 | 0 | 2 | 0 | 1 | 1 | 2 | 1 | 41 | 2 | 2 | 2 | 1 | 129 |
| 0 | 0 | 1 | 2 | 0 | 2 | 2 | 1 | 1 | 41 | 1 | 2 | 2 | 2 | 134 |
| 0 | 0 | 2 | 0 | 0 | 1 | 2 | 0 | 0 | 48 | 2 | 1 | 1 | 0 | 134 |
| 0 | 1 | 0 | 2 | 0 | 2 | 1 | 0 | 2 | 48 | 0 | 1 | 1 | 2 | 136 |
| 0 | 2 | 0 | 0 | 0 | 0 | 1 | 0 | 1 | 51 | 0 | 2 | 2 | 0 | 136 |
| 1 | 0 | 0 | 2 | 0 | 0 | 2 | 0 | 1 | 51 | 1 | 2 | 1 | 0 | 146 |
| 2 | 0 | 0 | 0 | 0 | 1 | 1 | 0 | 0 | 60 | 2 | 1 | 2 | 2 | 146 |
| 1 | 1 | 0 | 1 | 3 | 2 | 2 | 0 | 2 | 60 | 1 | 1 | 2 | 0 | 148 |
| 2 | 2 | 0 | 1 | 3 | 1 | 1 | 2 | 2 | 66 | 2 | 2 | 1 | 2 | 148 |
| 1 | 0 | 1 | 1 | 5 | 2 | 2 | 1 | 0 | 66 | 0 | 1 | 2 | 1 | 153 |
| 2 | 0 | 2 | 1 | 5 | 0 | 1 | 2 | 0 | 68 | 0 | 2 | 1 | 1 | 153 |
| 1 | 0 | 2 | 2 | 10 | 0 | 2 | 1 | 2 | 68 | 1 | 0 | 1 | 2 | 160 |
| 2 | 0 | 1 | 0 | 10 | 1 | 2 | 1 | 1 | 73 | 2 | 0 | 2 | 0 | 160 |
| 1 | 2 | 0 | 2 | 12 | 2 | 1 | 2 | 1 | 73 | 1 | 0 | 2 | 1 | 165 |
| 2 | 1 | 0 | 0 | 12 | 1 | 0 | 2 | 0 | 80 | 2 | 0 | 1 | 1 | 165 |
| 1 | 0 | 0 | 1 | 15 | 2 | 0 | 1 | 2 | 80 | 0 | 0 | 1 | 0 | 170 |
| 2 | 0 | 0 | 1 | 15 | 0 | 0 | 1 | 1 | 85 | 0 | 0 | 2 | 2 | 170 |
| 0 | 1 | 1 | 1 | 17 | 0 | 0 | 2 | 1 | 85 | 1 | 1 | 0 | 2 | 192 |
| 0 | 2 | 2 | 1 | 17 | 1 | 0 | 1 | 0 | 90 | 2 | 2 | 0 | 0 | 192 |
| 1 | 1 | 1 | 0 | 22 | 2 | 0 | 2 | 2 | 90 | 1 | 2 | 0 | 1 | 195 |
| 2 | 2 | 2 | 2 | 22 | 1 | 2 | 2 | 1 | 97 | 2 | 1 | 0 | 1 | 195 |
| 1 | 2 | 2 | 0 | 24 | 2 | 1 | 1 | 1 | 97 | 0 | 1 | 0 | 0 | 204 |
| 2 | 1 | 1 | 2 | 24 | 0 | 1 | 1 | 0 | 102 | 0 | 2 | 0 | 2 | 204 |
| 0 | 1 | 2 | 2 | 34 | 0 | 2 | 2 | 2 | 102 | 1 | 0 | 0 | 0 | 240 |
| 0 | 2 | 1 | 0 | 34 | 1 | 1 | 1 | 2 | 104 | 2 | 0 | 0 | 2 | 240 |
| 1 | 2 | 1 | 2 | 36 | 2 | 2 | 2 | 0 | 104 | 0 | 0 | 0 | 1 | 255 |

By inspection of Table 3, it can be seen that the 81 combinations generated only 53 different automata.

The filtering process consists of a repetitive composition of the modulo operator in order to obtain more possible values of the coefficients and yet get as the output of such transformation, the values necessary to define the automata in study. This way, for example, in order to get ternary cellular automata, the last composition needs to be with modulo 3, instead of modulo 2 as for the binary case.

Special care has to be taken while applying the filtering process. The modulo operation composition has to be taken with respect to prime numbers. This way, one has to situate the final filter – which determines the possible outputs of the transformation – regarding the prime numbers greater than such number. By doing this, the chance of getting a multiple of the modulus of congruence is diminished.

In order to represent every binary automaton, since the combinations of the coefficients do not uniquely define each rule, a compact notation has to be introduced to better represent the filtering process. This way, let the iota-delta function be defined as follows:

$$\iota\delta_n^m(x) = \mod[\mod[\mod[...\mod[\mod[x; p_m]; p_{m-1}];...]; p_{j+1}]; p_j]; n], \qquad (18)$$
$$m \geq j; \quad m, n \in \mathbb{Z}_+; \quad x \in \mathbb{C}; \quad j = \pi(n) + 1; \quad \max(p_m) \leq \max\big(\mathrm{Re}(x); \mathrm{Im}(x)\big)$$

in which $m$ and $n$ are parameters of the iota-delta function, $p_m$ is the $m$-th prime number and $\pi(n)$ stands for the prime counting function which gives the number of primes less than or equal to $n$. Note that it is considered that $p_1 = 2$. The value of $n$ determines how many states the automata generated have. Thus, for binary automaton, $n = 2$, for ternary ones, $n = 3$, for quaternary, $n = 4$ and so on. Also, $0 \leq \max\big(\iota\delta_n^m(x)\big) < n$. A Mathematica code which readily implements Eq. (18) is:

```
iotadelta[m1_,n1_,x1_]:=
    Module[{v1=PrimePi[n1]+1,y=Mod[x1,Prime[m1]]},
    While[v1<=m1-1,y=Mod[y,Prime[m1-v1+PrimePi[n1]]];v1++];
    Mod[y,n1]]
```

Based on Eq.(15), the number of combinations allowed by means of the iota-delta function is $p_m^4$ and the non-redundant coefficients are $\alpha_j = \{r | r \leq p_m - 1, r \in \mathbb{Z}_+\}$; and $j = 1,...,4$.

By means of the iota-delta function, the filtering process is better represented. For example, Eq.(17) can be written in a compact way as:

$$C_k^{i+1} = \iota\delta_2^2\left(\alpha_1 C_{k-1}^i + \alpha_2 C_k^i + \alpha_3 C_{k+1}^i + \alpha_4\right) \quad (19)$$

In order to represent every binary cellular automata in the simplest way possible, one must determine which is the smallest value of *m* such that for *n* = 2 every binary rule is expressed. By means of experimentation, when *m* = 5, i.e. mod(mod(mod(mod(mod(*x*;11);7);5);3);2), every binary cellular automaton is described. This way, by means of the iota-delta function every binary cellular automaton is represented by a single rule applied to the whole cellular net. Table 4 gives one of the combinations which generate each of the 0 – 255 Wolfram's rules.

Table 4. 0-255 Cellular Automata Coefficients

| Coefficients | | | | RNf | Coefficients | | | | RNf | Coefficients | | | | RNf | Coefficients | | | | RNf |
|---|---|---|---|---|---|---|---|---|---|---|---|---|---|---|---|---|---|---|---|
| $\alpha_1$ | $\alpha_2$ | $\alpha_3$ | $\alpha_4$ | | $\alpha_1$ | $\alpha_2$ | $\alpha_3$ | $\alpha_4$ | | $\alpha_1$ | $\alpha_2$ | $\alpha_3$ | $\alpha_4$ | | $\alpha_1$ | $\alpha_2$ | $\alpha_3$ | $\alpha_4$ | |
| 0 | 0 | 0 | 0 | 0 | 1 | 1 | 7 | 2 | 64 | 1 | 1 | 1 | 9 | 128 | 1 | 1 | 0 | 2 | 192 |
| 1 | 1 | 1 | 8 | 1 | 1 | 1 | 3 | 6 | 65 | 1 | 1 | 1 | 1 | 129 | 1 | 1 | 4 | 6 | 193 |
| 1 | 1 | 3 | 9 | 2 | 1 | 1 | 2 | 10 | 66 | 1 | 1 | 6 | 9 | 130 | 1 | 1 | 2 | 2 | 194 |
| 1 | 1 | 0 | 1 | 3 | 1 | 1 | 2 | 6 | 67 | 1 | 1 | 3 | 1 | 131 | 1 | 1 | 0 | 4 | 195 |
| 1 | 3 | 1 | 9 | 4 | 0 | 1 | 2 | 0 | 68 | 1 | 2 | 7 | 2 | 132 | 2 | 1 | 4 | 5 | 196 |
| 1 | 0 | 1 | 1 | 5 | 2 | 3 | 10 | 1 | 69 | 1 | 2 | 3 | 6 | 133 | 2 | 3 | 6 | 1 | 197 |
| 1 | 2 | 6 | 2 | 6 | 2 | 1 | 3 | 5 | 70 | 1 | 2 | 2 | 10 | 134 | 2 | 1 | 7 | 5 | 198 |
| 1 | 2 | 2 | 6 | 7 | 2 | 3 | 7 | 1 | 71 | 1 | 2 | 6 | 6 | 135 | 2 | 5 | 7 | 1 | 199 |
| 1 | 3 | 10 | 10 | 8 | 1 | 2 | 1 | 9 | 72 | 0 | 1 | 1 | 2 | 136 | 2 | 1 | 5 | 9 | 200 |
| 1 | 2 | 2 | 8 | 9 | 1 | 2 | 1 | 1 | 73 | 2 | 1 | 6 | 8 | 137 | 2 | 1 | 2 | 1 | 201 |
| 1 | 0 | 2 | 2 | 10 | 1 | 2 | 6 | 9 | 74 | 2 | 3 | 1 | 0 | 138 | 2 | 3 | 5 | 7 | 202 |
| 1 | 2 | 7 | 8 | 11 | 1 | 2 | 3 | 1 | 75 | 2 | 3 | 4 | 8 | 139 | 2 | 5 | 4 | 8 | 203 |
| 1 | 2 | 0 | 2 | 12 | 2 | 1 | 2 | 5 | 76 | 2 | 1 | 3 | 0 | 140 | 0 | 1 | 0 | 0 | 204 |
| 1 | 2 | 4 | 6 | 13 | 2 | 3 | 4 | 1 | 77 | 2 | 4 | 3 | 8 | 141 | 2 | 3 | 2 | 1 | 205 |
| 1 | 2 | 2 | 2 | 14 | 2 | 4 | 8 | 0 | 78 | 2 | 3 | 3 | 9 | 142 | 2 | 3 | 9 | 3 | 206 |
| 1 | 0 | 0 | 1 | 15 | 2 | 7 | 4 | 8 | 79 | 2 | 7 | 7 | 1 | 143 | 2 | 3 | 0 | 1 | 207 |
| 1 | 3 | 3 | 7 | 16 | 1 | 0 | 2 | 0 | 80 | 1 | 2 | 2 | 7 | 144 | 1 | 2 | 4 | 5 | 208 |
| 0 | 1 | 1 | 1 | 17 | 2 | 4 | 1 | 6 | 81 | 2 | 1 | 3 | 6 | 145 | 2 | 4 | 3 | 6 | 209 |
| 1 | 2 | 1 | 7 | 18 | 1 | 2 | 3 | 5 | 82 | 1 | 2 | 1 | 0 | 146 | 1 | 2 | 7 | 5 | 210 |
| 2 | 1 | 2 | 6 | 19 | 2 | 4 | 6 | 6 | 83 | 2 | 1 | 6 | 6 | 147 | 2 | 9 | 8 | 4 | 211 |
| 1 | 1 | 2 | 7 | 20 | 2 | 2 | 1 | 2 | 84 | 1 | 1 | 2 | 0 | 148 | 2 | 2 | 9 | 2 | 212 |
| 2 | 2 | 1 | 6 | 21 | 0 | 0 | 1 | 1 | 85 | 2 | 2 | 5 | 6 | 149 | 2 | 2 | 7 | 4 | 213 |
| 1 | 1 | 1 | 0 | 22 | 2 | 2 | 10 | 2 | 86 | 1 | 4 | 1 | 0 | 150 | 2 | 2 | 6 | 2 | 214 |
| 2 | 2 | 6 | 6 | 23 | 2 | 2 | 8 | 4 | 87 | 2 | 2 | 2 | 6 | 151 | 2 | 2 | 4 | 4 | 215 |
| 1 | 2 | 2 | 0 | 24 | 1 | 2 | 5 | 5 | 88 | 2 | 1 | 1 | 2 | 152 | 2 | 3 | 4 | 10 | 216 |
| 2 | 1 | 1 | 6 | 25 | 2 | 7 | 10 | 6 | 89 | 0 | 1 | 1 | 4 | 153 | 2 | 4 | 5 | 6 | 217 |
| 1 | 2 | 4 | 0 | 26 | 1 | 0 | 1 | 0 | 90 | 2 | 3 | 10 | 2 | 154 | 2 | 3 | 2 | 10 | 218 |
| 2 | 7 | 8 | 4 | 27 | 2 | 4 | 2 | 6 | 91 | 2 | 3 | 6 | 6 | 155 | 2 | 7 | 2 | 6 | 219 |
| 1 | 3 | 4 | 5 | 28 | 2 | 5 | 4 | 10 | 92 | 2 | 6 | 7 | 2 | 156 | 2 | 4 | 9 | 2 | 220 |
| 2 | 8 | 7 | 4 | 29 | 2 | 9 | 8 | 6 | 93 | 2 | 4 | 7 | 4 | 157 | 0 | 2 | 4 | 6 | 221 |
| 1 | 4 | 4 | 0 | 30 | 2 | 4 | 6 | 2 | 94 | 2 | 9 | 9 | 10 | 158 | 2 | 5 | 2 | 10 | 222 |
| 2 | 4 | 4 | 4 | 31 | 2 | 0 | 2 | 6 | 95 | 4 | 2 | 2 | 4 | 159 | 2 | 9 | 2 | 6 | 223 |
| 1 | 3 | 8 | 10 | 32 | 1 | 2 | 2 | 9 | 96 | 1 | 0 | 1 | 2 | 160 | 1 | 2 | 5 | 9 | 224 |
| 1 | 2 | 3 | 8 | 33 | 1 | 2 | 6 | 1 | 97 | 1 | 2 | 6 | 8 | 161 | 1 | 2 | 2 | 1 | 225 |
| 0 | 1 | 2 | 2 | 34 | 2 | 1 | 6 | 9 | 98 | 2 | 4 | 1 | 5 | 162 | 2 | 4 | 8 | 9 | 226 |
| 2 | 1 | 7 | 8 | 35 | 2 | 1 | 3 | 1 | 99 | 2 | 4 | 5 | 1 | 163 | 2 | 9 | 3 | 1 | 227 |
| 1 | 2 | 1 | 10 | 36 | 2 | 1 | 5 | 5 | 100 | 1 | 2 | 1 | 2 | 164 | 2 | 3 | 7 | 3 | 228 |
| 1 | 2 | 1 | 6 | 37 | 2 | 3 | 1 | 1 | 101 | 1 | 0 | 1 | 4 | 165 | 2 | 3 | 9 | 1 | 229 |
| 2 | 1 | 4 | 0 | 38 | 0 | 1 | 1 | 0 | 102 | 2 | 7 | 1 | 5 | 166 | 2 | 4 | 6 | 0 | 230 |
| 2 | 7 | 3 | 1 | 39 | 2 | 3 | 3 | 1 | 103 | 2 | 4 | 9 | 8 | 167 | 2 | 7 | 9 | 8 | 231 |
| 1 | 1 | 2 | 9 | 40 | 1 | 1 | 1 | 2 | 104 | 2 | 2 | 4 | 9 | 168 | 2 | 2 | 2 | 0 | 232 |
| 1 | 1 | 2 | 1 | 41 | 1 | 1 | 1 | 4 | 105 | 2 | 2 | 1 | 1 | 169 | 2 | 2 | 5 | 8 | 233 |
| 2 | 2 | 1 | 5 | 42 | 2 | 2 | 6 | 0 | 106 | 0 | 0 | 1 | 0 | 170 | 2 | 2 | 4 | 0 | 234 |
| 2 | 2 | 5 | 1 | 43 | 2 | 2 | 9 | 8 | 107 | 2 | 2 | 3 | 1 | 171 | 2 | 2 | 7 | 8 | 235 |
| 1 | 3 | 5 | 9 | 44 | 2 | 6 | 2 | 0 | 108 | 2 | 5 | 3 | 7 | 172 | 2 | 4 | 2 | 0 | 236 |
| 1 | 3 | 2 | 1 | 45 | 2 | 9 | 2 | 8 | 109 | 2 | 4 | 5 | 8 | 173 | 2 | 5 | 9 | 1 | 237 |
| 2 | 8 | 4 | 0 | 46 | 2 | 4 | 4 | 0 | 110 | 2 | 9 | 3 | 3 | 174 | 0 | 2 | 2 | 2 | 238 |
| 2 | 4 | 7 | 8 | 47 | 4 | 2 | 9 | 6 | 111 | 2 | 0 | 3 | 1 | 175 | 2 | 9 | 9 | 8 | 239 |
| 1 | 2 | 0 | 0 | 48 | 1 | 2 | 2 | 5 | 112 | 1 | 2 | 3 | 0 | 176 | 1 | 0 | 0 | 0 | 240 |
| 2 | 1 | 4 | 6 | 49 | 2 | 4 | 4 | 6 | 113 | 2 | 3 | 4 | 6 | 177 | 3 | 2 | 2 | 1 | 241 |
| 2 | 1 | 2 | 2 | 50 | 2 | 3 | 5 | 10 | 114 | 2 | 4 | 7 | 10 | 178 | 3 | 2 | 9 | 3 | 242 |
| 0 | 1 | 0 | 1 | 51 | 2 | 7 | 9 | 6 | 115 | 2 | 7 | 2 | 4 | 179 | 2 | 4 | 0 | 6 | 243 |
| 1 | 3 | 2 | 5 | 52 | 2 | 4 | 8 | 2 | 116 | 1 | 3 | 6 | 5 | 180 | 3 | 7 | 4 | 5 | 244 |
| 2 | 4 | 6 | 4 | 53 | 2 | 9 | 7 | 6 | 117 | 2 | 8 | 9 | 4 | 181 | 2 | 0 | 4 | 6 | 245 |
| 4 | 1 | 4 | 0 | 54 | 2 | 5 | 5 | 10 | 118 | 2 | 6 | 2 | 2 | 182 | 2 | 4 | 4 | 2 | 246 |
| 2 | 8 | 2 | 4 | 55 | 0 | 2 | 2 | 6 | 119 | 2 | 4 | 2 | 4 | 183 | 2 | 9 | 9 | 6 | 247 |
| 1 | 3 | 7 | 7 | 56 | 1 | 3 | 3 | 0 | 120 | 2 | 4 | 3 | 10 | 184 | 3 | 7 | 7 | 9 | 248 |
| 2 | 10 | 7 | 6 | 57 | 2 | 6 | 9 | 4 | 121 | 2 | 5 | 4 | 6 | 185 | 2 | 4 | 7 | 6 | 249 |
| 2 | 4 | 5 | 10 | 58 | 2 | 8 | 2 | 2 | 122 | 2 | 9 | 4 | 2 | 186 | 2 | 0 | 2 | 2 | 250 |
| 2 | 8 | 9 | 6 | 59 | 2 | 4 | 9 | 6 | 123 | 0 | 2 | 3 | 1 | 187 | 2 | 9 | 2 | 4 | 251 |
| 1 | 1 | 0 | 0 | 60 | 2 | 2 | 8 | 2 | 124 | 2 | 2 | 3 | 10 | 188 | 2 | 2 | 0 | 2 | 252 |
| 2 | 2 | 4 | 6 | 61 | 2 | 2 | 6 | 4 | 125 | 2 | 2 | 7 | 6 | 189 | 2 | 2 | 9 | 4 | 253 |
| 2 | 2 | 7 | 10 | 62 | 2 | 2 | 4 | 2 | 126 | 2 | 2 | 5 | 10 | 190 | 2 | 2 | 2 | 2 | 254 |
| 2 | 2 | 0 | 6 | 63 | 2 | 2 | 2 | 4 | 127 | 2 | 2 | 9 | 6 | 191 | 0 | 0 | 0 | 1 | 255 |

In the following section a physical interpretation of the iota-delta function is given. Besides, a general rule for every cellular automaton is also conjectured to exist.

## Physical Interpretation of The Transformation and The Capital Iota-Delta Function

An immediate physical interpretation of the iota-delta function is that nature is structured in terms of prime numbers. Since cellular automata describe some of the most complex behaviors in nature and their representation is given in terms of the iota-delta function, the modular structure of the latter has to be related to the former.

The filtering process can be compared to a lens effect. The reality is defined by means of the modular composition of primes and, while observing natural phenomena, one has to choose which lens to use. By choosing the modulo 2 perspective, the observer is considering only the binary states, such as existence or non-existence, matter or void, up or down (following Ising model of spins). On the other hand, by choosing other modulus of congruence, more states are recognizable, thus more phenomena are described. This intuition has to be further developed, notwithstanding, the general idea is given in these few lines.

Since the most basic operations are sum, subtraction, multiplication and division, it is expected that cellular automata can be described by means of such operations. This way, instead of considering the linear combination of cells, one has to consider the situation in which product is also allowed. Both difference and division follow from sum and product, this way, let the combination of powers, sums and products of cells be taken into account. The capital iota-delta function can be defined:

$$_\omega^\gamma \mathrm{I}\Delta_n^m \left[ C_k^i \left| \begin{matrix} (\beta_1,\mu_1),(\beta_2,\mu_2),...,(\beta_\omega,\mu_\omega) \\ [\![ u_1,u_2,...,u_\omega, \alpha_{u_1,u_2,...,u_\omega} ]\!] \end{matrix} \right. \right] = \iota\delta_n^m \left( \sum_{\lambda=0}^{\gamma} \left[ \sum_{u_1+u_2+...+u_\omega=\lambda} \alpha_{u_1,u_2,...,u_\omega} \prod_{1\le c\le\omega} \left( C_{k+\beta_c}^{i+\mu_c} \right)^{u_c} \right] \right) \quad (20)$$

in which:

1)  The second summation is taken over all combinations of nonnegative integer indices $u_1$ through $u_\omega$ such that the sum of all $u_j$ is $\lambda$.

2)  $\omega$ is the number of terms which are being combined. In other words, it represents how many cells the value of another given cell depends on. In Eq.(19), for example, $\omega =$ 3 since the value of $C_k^{i+1}$ depends of the value of three other cells, namely $C_{k-1}^i$, $C_k^i$ and $C_{k+1}^i$.

3) $\gamma$ is the greatest power of any combined term. In Eq. (19), $\gamma = 1$ since there are no terms on the right-hand side which are raised to powers greater than one.

4) *m* and *n* are the parameters of the iota-delta function.

5) The pairs $(\beta_c, \mu_c)$, $c = 1\ldots \omega$, are functional parameters which locate the combined terms with respect to $C_k^i$. For example, for $C_{k-1}^i$, the pair is (-1,0).

6) $\alpha_{u_1, u_2, \ldots, u_\omega}$ are the coefficients of the final terms for each set of values $u_1, u_2, \ldots, u_\omega$. In fact there is a total of $(\gamma+\omega)!/[\omega!\gamma!]$ of such coefficients. Besides, their values are $\alpha_{u_1, u_2, \ldots, u_\omega} = \{r | r \leq p_m - 1, r \in \mathbb{Z}_+\}$. The notation $[\![.]\!]$ stands for a matrix $(\gamma+\omega)!/[\omega!\gamma!]$ by $\omega+1$ whose lines are correspondent to each set of values $u_1, u_2, \ldots, u_\omega, \alpha_{u_1, u_2, \ldots, u_\omega}$. In Eq.(19), the correspondent lines are $(0,0,0,\alpha_{0,0,0} = \alpha_4)$, $(1,0,0,\alpha_{1,0,0} = \alpha_1)$, $(0,1,0,\alpha_{0,1,0} = \alpha_2)$, and $(0,0,1,\alpha_{0,0,1} = \alpha_3)$.

Based on the capital and the ordinary iota-delta functions, the general equation which represents every binary cellular automata is given as:

$$C_k^{i+1} = {}_3^1I\Delta_2^5 \left[ C_k^i \begin{Vmatrix} (-1,0), (0,0), (1,0) \\ \begin{matrix} 0 & 0 & 0 & \alpha_{0,0,0} \\ 1 & 0 & 0 & \alpha_{1,0,0} \\ 0 & 1 & 0 & \alpha_{0,1,0} \\ 0 & 0 & 1 & \alpha_{0,0,1} \end{matrix} \end{Vmatrix} \right] \quad (21)$$

$$= \iota\delta_2^5 \left( \alpha_{0,0,0} + \alpha_{1,0,0}C_{k-1}^i + \alpha_{0,1,0}C_k^i + \alpha_{0,0,1}C_{k+1}^i \right)$$

A few properties of the capital iota-delta function are given below:

1) Translation in *i* and *k*: If the cell $C_k^{i+1}$ is given as the capital iota-delta function in Eq.(20), then any cell $C_{k+\theta}^{i+\varphi}$ can be readily obtained as:

$$C_{k+\theta}^{i+\varphi} = {}_\omega^\gamma I\Delta_n^m \left[ C_k^i \begin{Vmatrix} (\beta_1+\theta, \mu_1+\varphi-1), (\beta_2+\theta, \mu_2+\varphi-1), \ldots, (\beta_\omega+\theta, \mu_\omega+\varphi-1) \\ [\![u_1, u_2, \ldots, u_\omega, \alpha_{u_1, u_2, \ldots, u_\omega}]\!] \end{Vmatrix} \right] \quad (22)$$

Proof: Let $C_k^{i+1}$ be defined as:

$$C_k^{i+1} = {}_\omega^\gamma I\Delta_n^m \left[ C_k^i \begin{Vmatrix} (\beta_1, \mu_1), (\beta_2, \mu_2), \ldots, (\beta_\omega, \mu_\omega) \\ [\![u_1, u_2, \ldots, u_\omega, \alpha_{u_1, u_2, \ldots, u_\omega}]\!] \end{Vmatrix} \right] \quad (23)$$

By making the index transformation $i = i + \varphi - 1$ and $k = k + \theta$, Eq.(22) follows.

2) Special case: when the coefficients $\alpha_{u_1,u_2,...,u_\omega} = \dfrac{\gamma!}{u_1!u_2!...u_\omega!}$, the capital iota-delta function reduces to:

$$_\omega^\gamma I\Delta_n^m \left[ C_k^i \left\| \begin{matrix} (\beta_1,\mu_1),(\beta_2,\mu_2),...,(\beta_\omega,\mu_\omega) \\ u_1,u_2,...,u_\omega, \dfrac{\gamma!}{u_1!u_2!...u_\omega!} \end{matrix} \right\| \right] = \iota\delta_n^m \left( \dfrac{\left(\sum_{c=1}^\omega C_{k+\beta_c}^{i+\mu_c}\right)^{\gamma+1} - 1}{\sum_{c=1}^\omega C_{k+\beta_c}^{i+\mu_c} - 1} \right) \quad (24)$$

Proof: Consider the multinomial theorem stated as Hosch (2011):

$$(x_1 + x_2 + ... + x_\omega)^\lambda = \sum_{u_1+u_2+...+u_\omega=\lambda} \dfrac{\gamma!}{u_1!u_2!...u_\omega!} \prod_{1 \leq c \leq \omega} (x_c)^{u_c} \quad (25)$$

Let $x_c = C_{k+\beta_c}^{i+\mu_c}$, $c = 1, 2, ..., \omega$. This way, when $\alpha_{u_1,u_2,...,u_\omega} = \dfrac{\gamma!}{u_1!u_2!...u_\omega!}$, Eq.(20) provides:

$$_\omega^\gamma I\Delta_n^m \left[ C_k^i \left\| \begin{matrix} (\beta_1,\mu_1),(\beta_2,\mu_2),...,(\beta_\omega,\mu_\omega) \\ u_1,u_2,...,u_\omega, \dfrac{\gamma!}{u_1!u_2!...u_\omega!} \end{matrix} \right\| \right] = \iota\delta_n^m \left( \sum_{\lambda=0}^\gamma \left[ \sum_{c=1}^\omega C_{k+\beta_c}^{i+\mu_c} \right]^\lambda \right)$$

$$= \iota\delta_n^m \left( \dfrac{\left(\sum_{c=1}^\omega C_{k+\beta_c}^{i+\mu_c}\right)^{\gamma+1} - 1}{\sum_{c=1}^\omega C_{k+\beta_c}^{i+\mu_c} - 1} \right) \quad (26)$$

Thus, Eq.(24) holds true.

# Application of The Transformation to Determine The Value of Any Cell in the Cellular Automaton Net

One of the most amazing patterns obtained by Wolfram (2002) was rule 30. In his book, Wolfram showed great concernment about the chaoticity and randomness of such rule. By means of the iota-delta function it is possible to answer some questions Wolfram made. In short, not only to rule 30, but to every cellular automata, the iota-delta function allows one to obtain explicitly the value of any cell in the cellular automaton net.

At first, by means of Table 4, a possible representation of rule 30 in terms of the capital iota-delta function is:

$$C_k^{i+1} = {}_3^1 I\Delta_2^5 \left[ C_k^i \begin{Vmatrix} (-1,0),(0,0),(1,0) \\ \begin{matrix} 0 & 0 & 0 & 0 \\ 1 & 0 & 0 & 1 \\ 0 & 1 & 0 & 4 \\ 0 & 0 & 1 & 4 \end{matrix} \end{Vmatrix} \right] \tag{27}$$

On the other hand, Eq.(27) can be represented in terms of the iota-delta function as:

$$C_k^{i+1} = \iota\delta_2^5 \left( C_{k-1}^i + 4C_k^i + 4C_{k+1}^i \right) \tag{28}$$

Note that Eq.(28) is valid for every cell in the cellular automaton net, this way, by the translation property of both the capital and the ordinary iota-delta function, it is true that:

$$\begin{aligned} C_{k-1}^i &= \iota\delta_2^5 \left( C_{k-2}^{i-1} + 4C_{k-1}^{i-1} + 4C_k^{i-1} \right) \\ C_k^i &= \iota\delta_2^5 \left( C_{k-1}^{i-1} + 4C_k^{i-1} + 4C_{k+1}^{i-1} \right) \\ C_{k+1}^i &= \iota\delta_2^5 \left( C_k^{i-1} + 4C_{k+1}^{i-1} + 4C_{k+2}^{i-1} \right) \end{aligned} \tag{29}$$

By Eqs.(28) and (29):

$$\begin{aligned} C_k^{i+1} = \iota\delta_2^5 \Big( &\iota\delta_2^5 \left( C_{k-2}^{i-1} + 4C_{k-1}^{i-1} + 4C_k^{i-1} \right) + 4\iota\delta_2^5 \left( C_{k-1}^{i-1} + 4C_k^{i-1} + 4C_{k+1}^{i-1} \right) + \\ &+ 4\iota\delta_2^5 \left( C_k^{i-1} + 4C_{k+1}^{i-1} + 4C_{k+2}^{i-1} \right) \Big) \end{aligned} \tag{30}$$

It can be seen that while Eq.(28) gives the value of $C_k^{i+1}$ in terms of the past step in $i$, Eq.(30) gives the same value in terms of two past steps in $i$. If one wants to get the value of $C_k^{i+1}$ in terms of three past steps in $i$, Eq.(30) has to be composed with an equation of the type of Eq.(28). On the other hand, if one composes Eq.(30) with itself, the value is going to be obtained in terms of 4 past steps in $i$. This process can be repeated in order to obtain the value of $C_k^{i+1}$ in terms of $h$ past steps.

By setting the initial condition as $C_0^0 = 1; C_{k \neq 0}^0 = 0$, which is the standard initial conditions for cellular automata, by means of the composition described above, the value of $C_k^h$ is easily obtained. For example, the term $C_0^h$, based on Eqs.(28) and (30), is given as:

$$C_0^h = \begin{cases} \iota\delta_2^5(4) = 1, h = 1 \\ \iota\delta_2^5 \left( \iota\delta_2^5(4) + 4\iota\delta_2^5(4) + 4\iota\delta_2^5(1) \right) = \iota\delta_2^5(1+4+4) = 0, h = 2 \end{cases} \tag{31}$$

which agrees with the rule 30 representation. By means of the capital iota-delta function, the function composition described above is given as:

$$
\begin{aligned}
C_{k+\theta}^{i+\varphi} &= {}^{\gamma}\mathrm{I}\Delta_n^m \left[ C_k^i \left| \begin{array}{l} (\beta_1+\theta,\mu_1+\varphi-1),(\beta_2+\theta,\mu_2+\varphi-1),...,(\beta_\omega+\theta,\mu_\omega+\varphi-1) \\ [u_1,u_2,...,u_\omega,\alpha_{u_1,u_2,...,u_\omega}] \end{array} \right. \right] \\
&= \iota\delta_n^m \left( \sum_{\lambda=0}^{\gamma} \left[ \sum_{u_1+u_2+...+u_\omega=\lambda} \alpha_{u_1,u_2,...,u_\omega} \prod_{1\le c\le\omega} \left( C_{k+\beta_c+\theta}^{i+\mu_c+\varphi-1} \right)^{u_c} \right] \right) \\
&= \iota\delta_n^m \left( \sum_{\lambda=0}^{\gamma} \left[ \sum_{u_1+u_2+...+u_\omega=\lambda} \alpha_{u_1,u_2,...,u_\omega} \prod_{1\le c\le\omega} \left( {}^{\gamma}\mathrm{I}\Delta_n^m \left[ C_k^i \left| \begin{array}{l} (\beta_1+\beta_c+\theta,\mu_1+\mu_c+\varphi-2),...,(\beta_\omega+\beta_c+\theta,\mu_\omega+\mu_c+\varphi-2) \\ [u_1,u_2,...,u_\omega,\alpha_{u_1,u_2,...,u_\omega}] \end{array} \right. \right] \right)^{u_c} \right] \right)
\end{aligned}
\tag{32}
$$

Note that Eq.(32) is a single composition. In order to give values with respect to more previous steps, one has to use the ordinary iota-delta function to represent the capital iota-delta function and then apply the composition.

Thus, every cell in the cellular automaton net can be explicitly determined based on the initial conditions and both the ordinary and capital iota-delta function.

# Quantitative Interpretation of Cellular Automata By Means of the Iota-Delta Function

In the previous section it has been demonstrated that every 0-255 cellular automata can be represented in terms of the iota-delta function as Eq. (21). In the present section, a quantitative interpretation of such relation is deduced by comparing Eq. (21) to the Finite Difference Method (FDM).

As widely known, FDM is a numerical method which turns differential equations into difference equations, making possible to solution of the former by solving a system of the latter. The methodology to be presented is applicable to every partial differential equation (PDE), on the other hand, in order to better explain the procedure, the advective-dispersive equation which describes, for example, the solute flow in a porous medium is taken into account.

This way, let the equation which describes the concentration $c(x,t)$ of a given solute flowing in a porous medium as (Najafi e Hajinezhad, 2008):

$$\frac{\partial c}{\partial t} = D_x \frac{\partial^2 c}{\partial x^2} - v_x \frac{\partial c}{\partial x} \tag{33}$$

in which $D_x$ [L$^2$/T] is the hydrodynamic dispersivity of the medium and $v_x$ [L/T] is the mean velocity of the interstitial fluid.

The readership is probably familiarized with the concepts behind FDM, this way, basic definitions will be omitted. The latter can be further investigated in (LeVeque, 2007). Let one use the forward difference in space and time for the first order derivative. Also, the central difference for the second order derivative in space shall be used. This way, by applying FDM to Eq. (33) considering a structured mesh whose lengths in space and time are $\Delta t$ and $\Delta x$, respectively, the latter equation turns to:

$$\frac{c(x,t+\Delta t)-c(x,t)}{\Delta t} = D_x \frac{c(x+\Delta x,t)-2c(x,t)+c(x-\Delta x,t)}{\Delta x^2} - v_x \frac{c(x+\Delta x,t)-c(x,t)}{\Delta x} \quad (34)$$

In order to simplify notation and also clarify the link between FDM and cellular automata, let one denote $k$ as the position $x$ in space and $i$ the position $t$ in time. It is clear that based on the mesh lengths described above, the position $k+1$ is equivalent to $x + \Delta x$ in the "real" space mesh and the position $i+1$ is equivalent to $t + \Delta t$ in the "real" time mesh. Also, let the following substitutions take place:

$$\frac{D_x \Delta t}{\Delta x^2} = N \quad (35)$$

$$\frac{v_x \Delta t}{\Delta x} = C_r \quad (36)$$

in which $N$ and $C_r$ are the Neumann and Courant numbers, respectively. This way, Eq. (34) turns to:

$$c_k^{i+1} = (N - C_r) c_{k+1}^i + (1 + C_r - 2N) c_k^i + N c_{k-1}^i \quad (37)$$

It has been shown by Ataie-Ashtiani et al. (1999) that the FDM scheme used above is convergent if:

$$2N + C_r \leq 1 \quad (38)$$

The similarity between Eqs. (21) and (37) is remarkable. The main question is how to directly relate one to the other. An immediate alternative is linear scaling the iota-delta function in order to obtain Eq. (37). This way, consider a non-null scaling constant $S$ multiplied to the right-hand side of Eq. (21). When $S = 1$ Eq. (21) is recovered. The new rule takes the form:

$$C_k^{i+1} = S \iota \delta_2^5 \left( \alpha_{0,0,0} + \alpha_{1,0,0} C_{k-1}^i + \alpha_{0,1,0} C_k^i + \alpha_{0,0,1} C_{k+1}^i \right) \quad (39)$$

Consider the evolution of Eq. (39) for a unitary initial condition presented in Figure 5.

|   |   | 1 |   |   |
|---|---|---|---|---|
|   | $V_1$ | $V_2$ | $V_3$ |   |
| $V_4$ | $V_5$ | $V_6$ | $V_7$ | $V_8$ |

Figure 5. Evolution of Eq. (39)

Note that the values of $V_b$, $b = 1, 2,…,8$ are given in Table 5.

Table 5. Coefficients Concerning Figure 5.

| Values of $V_b$ | Correspondent iota-delta values |
|---|---|
| $V_1$ | $S\iota\delta_2^5(\alpha_{0,0,1} + \alpha_{0,0,0})$ |
| $V_2$ | $S\iota\delta_2^5(\alpha_{0,1,0} + \alpha_{0,0,0})$ |
| $V_3$ | $S\iota\delta_2^5(\alpha_{1,0,0} + \alpha_{0,0,0})$ |
| $V_4$ | $S\iota\delta_2^5(\alpha_{0,0,1}S\iota\delta_2^5(\alpha_{0,0,1} + \alpha_{0,0,0}) + \alpha_{0,0,0})$ |
| $V_5$ | $S\iota\delta_2^5(\alpha_{0,1,0}S\iota\delta_2^5(\alpha_{0,0,1} + \alpha_{0,0,0}) + \alpha_{0,0,1}S\iota\delta_2^5(\alpha_{0,1,0} + \alpha_{0,0,0}) + \alpha_{0,0,0})$ |
| $V_6$ | $S\iota\delta_2^5(\alpha_{1,0,0}S\iota\delta_2^5(\alpha_{0,0,1} + \alpha_{0,0,0}) + \alpha_{0,1,0}S\iota\delta_2^5(\alpha_{0,1,0} + \alpha_{0,0,0}) + \alpha_{0,0,1}S\iota\delta_2^5(\alpha_{1,0,0} + \alpha_{0,0,0}) + \alpha_{0,0,0})$ |
| $V_7$ | $S\iota\delta_2^5(\alpha_{1,0,0}S\iota\delta_2^5(\alpha_{0,1,0} + \alpha_{0,0,0}) + \alpha_{0,1,0}S\iota\delta_2^5(\alpha_{1,0,0} + \alpha_{0,0,0}) + \alpha_{0,0,0})$ |
| $V_8$ | $S\iota\delta_2^5(\alpha_{1,0,0}S\iota\delta_2^5(\alpha_{1,0,0} + \alpha_{0,0,0}) + \alpha_{0,0,0})$ |

The coefficients shown in Table 5 were obtained by simply applying the general rule on Eq. (39).

In order to determine which the value of $S$ is, one shall compare the evolution of Eq. (39) to the evolution of the FDM scheme in Eq. (37). The evolution process presented in Figure 5 is also applicable to the FDM scheme, on the other hand, the new values of $V_b$, $b = 1, 2,…, 8$ are shown in Table 6.

First and foremost, in order be able to compare the values of $V_b$, it is important to notice that the FDM scheme adopted depends only on a linear combination of the values of the variable of interest in a previous step. This way, $\alpha_{0,0,0} = 0$. In order to turn the values $V_b$ concerning the cellular automata (CA) approach independent of the iota delta-function, the argument of the iota-delta function in Eq. (39) has to be less than 2. This comes from the definition in Eq. (9) taking $s = n = 2$, i. e., for the 0-255 cellular automata. From Table 5, since one is dealing with binary cellular automata,

$$\max(V_b) = S \max\left(\iota\delta_2^5(x)\right) = S, \quad b = 1, 2, 3.$$ In special, if the maximum of the argument of the iota-delta function which defines $V_6$ is less than 2, every other argument will also be. This way:

$$\alpha_{1,0,0} C_{k-1}^i + \alpha_{0,1,0} C_k^i + \alpha_{0,0,1} C_{k+1}^i \leq \alpha_{1,0,0} S + \alpha_{0,1,0} S + \alpha_{0,0,1} S \tag{40}$$

Table 6. Evolution of the FDM scheme in Eq. (37).

| Values of $V_b$ | Correspondent FDM scheme values |
|---|---|
| $V_1$ | $N - C_r$ |
| $V_2$ | $1 + C_r - 2N$ |
| $V_3$ | $N$ |
| $V_4$ | $(N - C_r)^2$ |
| $V_5$ | $2(N - C_r)(1 + C_r - 2N)$ |
| $V_6$ | $2N(N - C_r) + (1 + C_r - 2N)^2$ |
| $V_7$ | $2N(1 + C_r - 2N)$ |
| $V_8$ | $N^2$ |

In order to find the value of $S$, one must notice that the coefficients of the FDM scheme, in each line, always add up to one. This way, consider that the first line of the CA methodology add up to $\Omega$. Thus:

$$S\left[\iota\delta_2^5(\alpha_{0,0,1}) + \iota\delta_2^5(\alpha_{0,1,0}) + \iota\delta_2^5(\alpha_{1,0,0})\right] = \Omega \tag{41}$$

Assuming that exists $S$ such that each line of the CA will add up to the same $\Omega$, in order to normalize the lines obtained from the CA approach one has to divide each $V_b$ by the latter constant. Finally, from Eq. (41) and Tables 5 and 6, it is possible to obtain the following system of equations:

$$\begin{cases} N - C_r = \dfrac{S\iota\delta_2^5(\alpha_{0,0,1})}{\Omega} = \dfrac{\iota\delta_2^5(\alpha_{0,0,1})}{\iota\delta_2^5(\alpha_{0,0,1}) + \iota\delta_2^5(\alpha_{0,1,0}) + \iota\delta_2^5(\alpha_{1,0,0})} \\ 1 + C_r - 2N = \dfrac{S\iota\delta_2^5(\alpha_{0,1,0})}{\Omega} = \dfrac{\iota\delta_2^5(\alpha_{0,1,0})}{\iota\delta_2^5(\alpha_{0,0,1}) + \iota\delta_2^5(\alpha_{0,1,0}) + \iota\delta_2^5(\alpha_{1,0,0})} \\ N = \dfrac{S\iota\delta_2^5(\alpha_{1,0,0})}{\Omega} = \dfrac{\iota\delta_2^5(\alpha_{1,0,0})}{\iota\delta_2^5(\alpha_{0,0,1}) + \iota\delta_2^5(\alpha_{0,1,0}) + \iota\delta_2^5(\alpha_{1,0,0})} \end{cases} \tag{42}$$

which can be readily solved in order to obtain:

$$\begin{cases} C_r = \dfrac{\iota\delta_2^5(\alpha_{1,0,0}) - \iota\delta_2^5(\alpha_{0,0,1})}{\iota\delta_2^5(\alpha_{0,0,1}) + \iota\delta_2^5(\alpha_{0,1,0}) + \iota\delta_2^5(\alpha_{1,0,0})} \\ N = \dfrac{\iota\delta_2^5(\alpha_{1,0,0})}{\iota\delta_2^5(\alpha_{0,0,1}) + \iota\delta_2^5(\alpha_{0,1,0}) + \iota\delta_2^5(\alpha_{1,0,0})} \end{cases} \quad (43)$$

Since $\alpha_{0,0,1}, \alpha_{0,1,0}, \alpha_{1,0,0}$ are all positive integers, their iota-delta functions in Eq. (43) are either 0 or 1. This way, it is interesting to observe the possible values of the courant number in Eq. (43). This way:

a) $C_r > 0$, which is an advective-dispersive phenomenon From Eq. (43) it is easy to see that:

$$\iota\delta_2^5(\alpha_{1,0,0}) - \iota\delta_2^5(\alpha_{0,0,1}) > 0 \Rightarrow \iota\delta_2^5(\alpha_{1,0,0}) = 1; \quad \iota\delta_2^5(\alpha_{0,0,1}) = 0 \quad (44)$$

b) $C_r = 0$, which is a pure diffusion system, from Eq. (43) one shall get:

$$\iota\delta_2^5(\alpha_{1,0,0}) - \iota\delta_2^5(\alpha_{0,0,1}) = 0 \Rightarrow \iota\delta_2^5(\alpha_{1,0,0}) = \iota\delta_2^5(\alpha_{0,0,1}) \quad (45)$$

c) $C_r < 0$, which characterize an advective-dispersive phenomenon, from Eq.(43), finally:

$$\iota\delta_2^5(\alpha_{1,0,0}) - \iota\delta_2^5(\alpha_{0,0,1}) < 0 \Rightarrow \iota\delta_2^5(\alpha_{1,0,0}) = 0; \quad \iota\delta_2^5(\alpha_{0,0,1}) = 1 \quad (46)$$

Given the possible values of the courant number, it is possible to use Eq. (38) to check the convergence of the CA models hereby defined. This way, from Eqs. (38), (44), (45) and (46), one shall get:

a) $C_r > 0$:

$$2N + C_r = \dfrac{3\iota\delta_2^5(\alpha_{1,0,0}) - \iota\delta_2^5(\alpha_{0,0,1})}{\iota\delta_2^5(\alpha_{0,0,1}) + \iota\delta_2^5(\alpha_{0,1,0}) + \iota\delta_2^5(\alpha_{1,0,0})} = \dfrac{3}{1 + \iota\delta_2^5(\alpha_{0,1,0})} > 1 \quad (47)$$

thus, the CA is never convergent in this case.

b) $C_r = 0$:

$$\dfrac{3\iota\delta_2^5(\alpha_{1,0,0}) - \iota\delta_2^5(\alpha_{0,0,1})}{\iota\delta_2^5(\alpha_{0,0,1}) + \iota\delta_2^5(\alpha_{0,1,0}) + \iota\delta_2^5(\alpha_{1,0,0})} = \dfrac{2\iota\delta_2^5(\alpha_{1,0,0})}{\iota\delta_2^5(\alpha_{0,1,0}) + 2\iota\delta_2^5(\alpha_{1,0,0})} \geq 1 \quad (48)$$

which demonstrates that CA schemes for pure diffusive problems are always convergent and explicit.

c) $C_r < 0$, in this case, Eq. (38) is slightly modified to $|2N + C_r| < 1$, which leads to:

$$\frac{3\iota\delta_2^5(\alpha_{1,0,0}) - \iota\delta_2^5(\alpha_{0,0,1})}{\iota\delta_2^5(\alpha_{0,0,1}) + \iota\delta_2^5(\alpha_{0,1,0}) + \iota\delta_2^5(\alpha_{1,0,0})} = \frac{-1}{1 + \iota\delta_2^5(\alpha_{0,1,0})}; -1 \leq \frac{-1}{1 + \iota\delta_2^5(\alpha_{0,1,0})} < 1 \qquad (49)$$

which shows that CA schemes for advective-diffusive problems, when Courant number is negative, are always convergent and explicit.

Finally, if one imposes that the right-hand side of Eq. (40) is less than 2, so will the left-hand side be. It can be shown that only when the right-hand side of Eq. (40) is equal to one, the condition that each line of the CA add up to $\Omega$ is satisfied. Thus:

$$\alpha_{1,0,0}S + \alpha_{0,1,0}S + \alpha_{0,0,1}S = 1 \Rightarrow S = \frac{1}{\alpha_{1,0,0} + \alpha_{0,1,0} + \alpha_{0,0,1}} \qquad (50)$$

By means of the formulas above, CA schemes prove their value as convergent explicit methods to describe systems once only described by means of PDEs. Also, it is interesting to explore the diffusive case, when Courant number is zero. Note that FDM schemes demand the coefficients in Eq. (37) to be symmetric, i.e., both the first and last equal to $N$. On the other hand, the CA scheme only requires that the iota-delta of the coefficients to be symmetric. This way, a very interesting feature of the CA scheme shows up: the possibility to deal with normal and anomalous diffusion by means of a single formulation. This will be better explained in the following section.

## Normal Diffusion and Anomalous Diffusion by means of Cellular Automata

As stated above, for a FDM scheme, normal diffusion is obtained when the coefficients are symmetric with respect to the initial condition space row. On the other hand, when the iota-delta functions of the coefficients which define the automaton rule are symmetric with respect to the initial condition space row, not only normal diffusion but also anomalous behavior is easily seen. At first, a question which demands an answer is whether the CA approach for describing diffusive problems gives the same answer as the FDM scheme. In order to address this issue, one has to take into account two conditions, namely:

a) Do the coefficients of the CA methodology are symmetric with respect to the initial condition line:

In order to address this first condition, let one consider the coefficients $V_4$ and $V_8$. Based on the observation above, in order to have symmetry with respect to the initial condition line, one must have:

$$\frac{S}{\Omega}\iota\delta_2^5\left(\alpha_{0,0,1}S\iota\delta_2^5\left(\alpha_{0,0,1}\right)\right)=\frac{S}{\Omega}\iota\delta_2^5\left(\alpha_{1,0,0}S\iota\delta_2^5\left(\alpha_{1,0,0}\right)\right) \tag{51}$$

From the conditions in Eq. (45), since the inner iota-delta functions in Eq. (51) are the same, they are either both 0 or 1. When they are 0, it is clear that Eq. (51) is fully satisfied. On the other hand, when they are both one, the argument of the outer iota delta function becomes less than two since $S$ is the inverse of sum of the three coefficients, which finally takes the argument out of the iota-delta function and leads to $\alpha_{0,0,1}=\alpha_{1,0,0}$. It has been proved, this way, that in order to have symmetric diffusion with respect to the initial condition position, one must have, that both the coefficients and their iota-delta functions are equal. Now one has to investigate the second and most important condition, which follows.

  b) Do the outer coefficients of the CA methodology are related to the ones of the previous step? If yes, how is the relation?

It is clear from table 6 that the outer coefficients of the FDM scheme are related to the ones on the previous steps by a power function. In order to better develop this part of the paper, Neumann numbers will also be indexed in space and time, thus, the FDM shows that, for a diffusive case:

$$N_{-i}^i=\left(N_{-1}^1\right)^i;N_i^i=\left(N_1^1\right)^i,i\geq 1 \tag{52}$$

From Eq. (43), the correspondent Neumann numbers are $N_{-1}^1=N_1^1=S/\Omega$. CA methodology, on the other hand, shows that:

$$N_{-i}^i=\frac{S}{\Omega}\iota\delta_2^5\left(\alpha_{0,0,1}\Omega N_{-i+1}^{i-1}\right);N_i^i=\frac{S}{\Omega}\iota\delta_2^5\left(\alpha_{1,0,0}\Omega N_{i-1}^{i-1}\right);i\geq 2 \tag{53}$$

By means of Eq. (53) it is easy to prove that:

$$N_{-i}^i=\frac{S^i\left(\alpha_{0,0,1}\right)^{i-1}}{\Omega};N_i^i=\frac{S^i\left(\alpha_{1,0,0}\right)^{i-1}}{\Omega};i\geq 2 \tag{54}$$

There are three situations concerning Eqs. (52) and (54), namely:

b.1) Two-sided lighter-tailed anomalous diffusion when:

$$N_{-i}^i = \frac{S^i (\alpha_{0,0,1})^{i-1}}{\Omega} < \frac{S^i}{\Omega^i}; N_i^i = \frac{S^i (\alpha_{1,0,0})^{i-1}}{\Omega} < \frac{S^i}{\Omega^i}; i \geq 1 \tag{55a}$$

By lighter-tailed one shall consider the comparison with the standard diffusion hereby represented by the FDM approach. Note that Eq. (55) physically suggests that the last values of the CA scheme are lower than the ones of the FDM scheme, which implies a lighter-tailed distribution of concentrations in the former compared to the latter. From Eq. (55) one shall get:

$$\Omega \alpha_{0,0,1} < 1; \quad \Omega \alpha_{1,0,0} < 1 \tag{55b}$$

By means of Eq. (41), Eq. (55b) turns to:

$$\alpha_{0,0,1} \iota \delta_2^5 (\alpha_{0,1,0}) < \alpha_{1,0,0} + \alpha_{0,1,0} - \alpha_{0,0,1}; \quad \alpha_{1,0,0} \iota \delta_2^5 (\alpha_{0,1,0}) < \alpha_{0,0,1} + \alpha_{0,1,0} - \alpha_{1,0,0} \tag{56}$$

One has to pay close attention to the fact that Eq. (56) gives the relation between the coefficients in order to be configured a two-sided lighter-tailed anomalous diffusion. The one-sided case takes place when only one of the inequalities in Eq. (56) is satisfied. For the symmetric case, when $\alpha_{0,0,1} = \alpha_{1,0,0}$, Eq. (56) turns to:

$$\alpha_{0,0,1} \iota \delta_2^5 (\alpha_{0,1,0}) < \alpha_{0,1,0}; \quad \alpha_{1,0,0} \iota \delta_2^5 (\alpha_{0,1,0}) < \alpha_{0,1,0} \tag{57}$$

As an example, consider rule 150. From Table 4 and Eq. (21) its Capital iota delta representation is:

$$C_k^{i+1} = {}_3^1 I \Delta_2^5 \left[ C_k^i \left\| \begin{matrix} (-1,0),(0,0),(1,0) \\ \begin{bmatrix} 0 & 0 & 0 & 0 \\ 1 & 0 & 0 & 1 \\ 0 & 1 & 0 & 4 \\ 0 & 0 & 1 & 1 \end{bmatrix} \end{matrix} \right\| \right] \tag{58}$$

Also, it is true that Eq. (43) provides $N = 1/3$. For the rule in Eq. (58), both inequalities in Eq. (57) are satisfied, which characterizes a two-sided lighter-tailed concentration distribution, as it can be shown in Figure 6 in which a correspondent FDM with $N = 1/3$ is also plotted. Since the relation between the discretized time and space and the real ones is $t = i\Delta t$ and to $x = k\Delta x$, respectively, the lengths of the mesh are just scale factors, this way, Figure 6 has been plotted with the initial condition in the position $i = k = 0$ and the mesh lengths were taken as the unit.

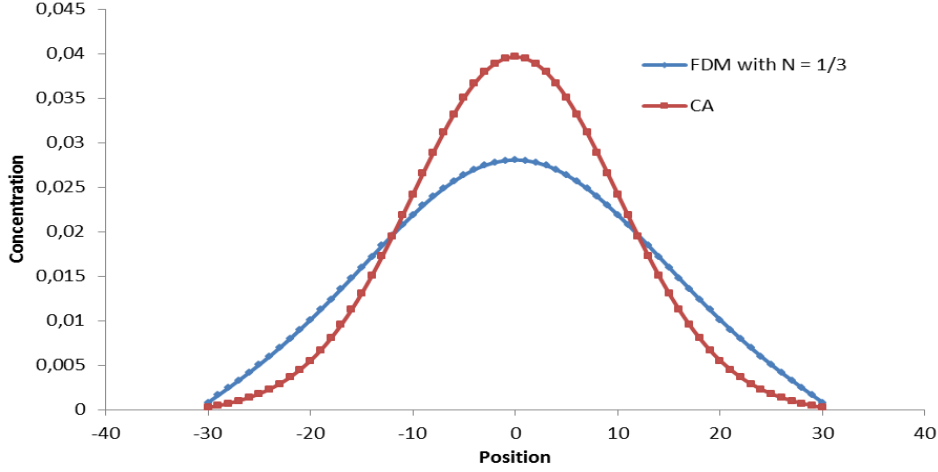

Figure 6. Comparison between the CA methodology for rule 150 and the Correspondent FDM scheme after 303 time steps.

By means of Eqs. (52) and (54), it is clear that Neumann number changes as the time steps evolve. In Eq. (35) the physical meaning of Neumann number has been given, which implies one to verify which physical parameter is changing over time. Considering that the net is not changing in time, from Eqs. (35), (52) and (54) one shall get that the diffusivity of the medium for a symmetric two-sided lighter-tailed diffusion is given as:

$$\left(\frac{D_x \Delta t}{\Delta x^2}\right)^i = \frac{S^i \left(\alpha_{0,0,1}\right)^{i-1}}{\Omega} \Rightarrow D_x = S\frac{\Delta x^2}{\Delta t}\frac{\left(\alpha_{0,0,1}\right)^{\frac{i-1}{i}}}{\sqrt[i]{\Omega}} \tag{59}$$

The discretization of the cellular net led to $t = i\Delta t$, thus Eq. (59) becomes:

$$D_x = S\frac{\Delta x^2}{\Delta t}\frac{\left(\alpha_{0,0,1}\right)^{\frac{t-\Delta t}{t}}}{\Omega^{\frac{\Delta t}{t}}}; \quad t \geq \Delta t \tag{60}$$

An amazing feature of Eq. (60) is the fact that the diffusivity varies following a power-law relation, which is the base of scaling properties in fractal structures. The comparison between the diffusivity of a CA and FDM schemes is given in Figure 7. The same considerations concerning the mesh lengths used to interpret Figure 6 are also applicable to Figure 7.

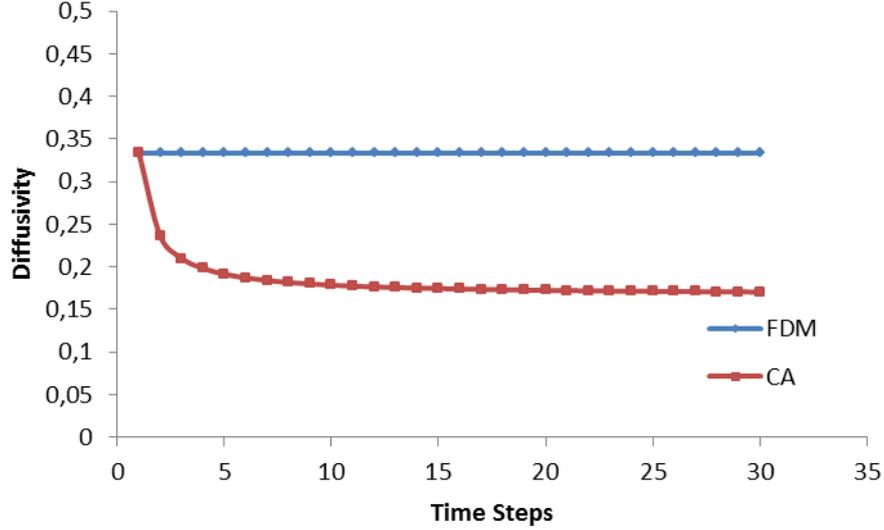

Figure 7. Comparison between the diffusivity of a CA methodology for rule 150 and the Correspondent FDM scheme.

From Eq. (59), when *i* tends to infinity, by means of standard limit application rules, it is easy to get:

$$D_x^{steady} = \lim_{i \to \infty} S \frac{\Delta x^2 (\alpha_{0,0,1})^{\frac{i-1}{i}}}{\Delta t \sqrt[i]{\Omega}} = S \frac{\Delta x^2 \alpha_{0,0,1}}{\Delta t} \qquad (61)$$

One shall note that using the limit in Eq. (61) in which a given value is taken to tend to infinity does not invalidate the discreetness of the approach since this type of limit only consider a large number of times steps, which is physically acceptable. The index steady has been added to the diffusivity in order to show that this value is obtained for large times. It is very interesting to observe that even if the CA scheme has constant coefficients in its generating rule, a time dependent behavior is seen for the diffusivity.

b.2) Two-sided light-tailed normal diffusion, or FDM approach:

This is the case when both the CA methodology and the FDM scheme give the same results. This way:

$$N_{-i}^i = \frac{S^i (\alpha_{0,0,1})^{i-1}}{\Omega} = \frac{S^i}{\Omega^i}; N_i^i = \frac{S^i (\alpha_{1,0,0})^{i-1}}{\Omega} = \frac{S^i}{\Omega^i}; i \geq 1 \qquad (62)$$

Following an argumentation similar to the one employed in item b.1, the following hold for the case when CA and FDM give the same results:

$$\Omega \alpha_{0,0,1} = 1; \quad \Omega \alpha_{1,0,0} = 1 \qquad (63)$$

$$\alpha_{0,0,1} \iota \delta_2^5 (\alpha_{0,1,0}) = \alpha_{1,0,0} + \alpha_{0,1,0} - \alpha_{0,0,1}; \quad \alpha_{1,0,0} \iota \delta_2^5 (\alpha_{0,1,0}) = \alpha_{0,0,1} + \alpha_{0,1,0} - \alpha_{1,0,0} \quad (64)$$

For the symmetric case:

$$\alpha_{0,0,1} \iota \delta_2^5 (\alpha_{0,1,0}) = \alpha_{0,1,0}; \quad \alpha_{1,0,0} \iota \delta_2^5 (\alpha_{0,1,0}) = \alpha_{0,1,0} \quad (65)$$

Also, from Eqs. (59) and (63), the diffusivity is given as:

$$\left( \frac{D_x \Delta t}{\Delta x^2} \right)^i = \frac{S^i (\alpha_{0,0,1})^i}{\Omega \alpha_{0,0,1}} \Rightarrow D_x = S \frac{\Delta x^2 \alpha_{0,0,1}}{\Delta t} \quad (66)$$

which is compatible with the normal diffusion situation in which the diffusivity is constant over time. Consider rule 22. Based on Table 4 and Eq. (21), rule 22 can be readily identified as:

$$C_k^{i+1} = {}_3^1 I \Delta_2^5 \left[ C_k^i \begin{Vmatrix} |(-1,0),(0,0),(1,0) \\ \begin{Vmatrix} 0 & 0 & 0 & 0 \\ 1 & 0 & 0 & 1 \\ 0 & 1 & 0 & 1 \\ 0 & 0 & 1 & 1 \end{Vmatrix} \end{Vmatrix} \right] \quad (67)$$

which also gives $N = 1/3$. Even Neumann numbers of the correspondent FDM schemes are the same for rules 150 and 22, only the latter is the correspondent FDM method itself, since Eqs. (62) to (66) are satisfied by the latter. There is no special physical meaning for the case b.2, thus no figures will be shown.

b.3) Two-sided heavier-tailed anomalous diffusion:

The term heavier-tailed is, as in b.1, with respect to the ordinary FDM scheme. By means of a similar approach as in b.1 and b.2, a two-sided heavier-tailed anomalous diffusion takes place when:

$$N_{-i}^i = \frac{S^i (\alpha_{0,0,1})^{i-1}}{\Omega} > \frac{S^i}{\Omega^i}; N_i^i = \frac{S^i (\alpha_{1,0,0})^{i-1}}{\Omega} > \frac{S^i}{\Omega^i}; i \geq 1 \quad (68)$$

Also:

$$\Omega \alpha_{0,0,1} > 1; \quad \Omega \alpha_{1,0,0} > 1 \quad (69)$$

$$\alpha_{0,0,1} \iota \delta_2^5 (\alpha_{0,1,0}) > \alpha_{1,0,0} + \alpha_{0,1,0} - \alpha_{0,0,1}; \quad \alpha_{1,0,0} \iota \delta_2^5 (\alpha_{0,1,0}) > \alpha_{0,0,1} + \alpha_{0,1,0} - \alpha_{1,0,0} \quad (70)$$

For the symmetric case:

$$\alpha_{0,0,1}\iota\delta_2^5(\alpha_{0,1,0}) > \alpha_{0,1,0}; \quad \alpha_{1,0,0}\iota\delta_2^5(\alpha_{0,1,0}) > \alpha_{0,1,0} \tag{71}$$

One shall note that Eqs. (59) to (61) are also valid for this case. As an example one shall take rule 54 and its capital iota delta representation. By Table 4 and Eq. (21):

$$C_k^{i+1} = \tfrac{1}{3}\mathrm{I}\Delta_2^5 \left[ C_k^i \left\| \begin{array}{c} |(-1,0),(0,0),(1,0)| \\ \begin{bmatrix} 0 & 0 & 0 & 0 \\ 1 & 0 & 0 & 4 \\ 0 & 1 & 0 & 1 \\ 0 & 0 & 1 & 4 \end{bmatrix} \end{array} \right\| \right] \tag{72}$$

which provides $N = 1/3$. Figures 8 and 9 show the behavior of rule 54 compared to the correspondent FDM scheme, based on the same premises used to investigate Figures 6 and 7.

In the case where heavier-tailed behavior is seen, diffusivity grows in time, achieving its steady state value given in Eq. (61).

An outstanding behavior can be seen by noticing that the steady state diffusivity is the diffusivity of another FDM scheme with different parameters, as the comparison of Eqs. (61) and (66) shows. This is a physically justified behavior since after a long period of time, the differences between anomalous and normal diffusion tend to zero as the concentration tends to be equally distributed all over the domain.

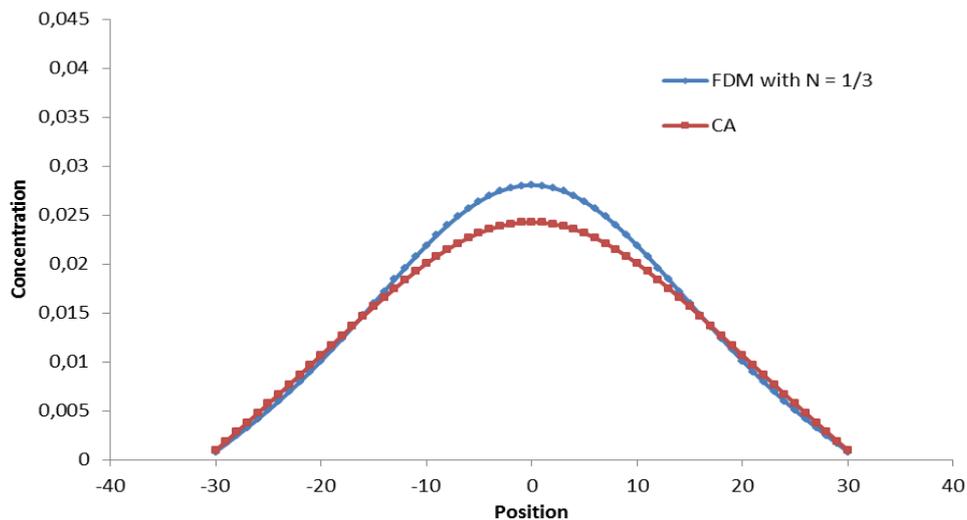

Figure 8. Comparison between the CA methodology for rule 54 and the Correspondent FDM scheme after 303 time steps.

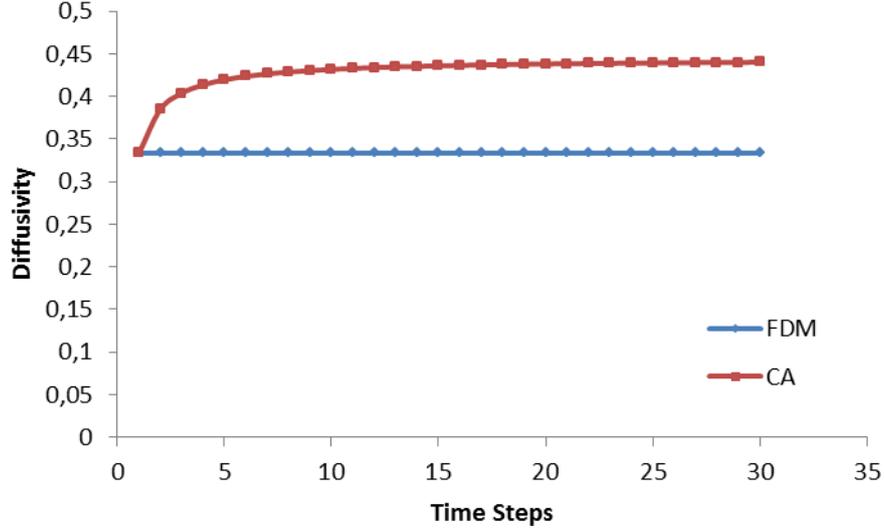

Figure 9. Comparison between the diffusivity of a CA methodology for rule 54 and the Correspondent FDM scheme.

Another interesting behavior can be seen when asymmetric diffusive rules are taken into account. In order to investigate this behavior, consider, for example, rule 146. From Table 4 and Eq. (21) its Capital iota delta representation is:

$$C_k^{i+1} = \tfrac{1}{3}\mathrm{I}\Delta_2^5 \left[ C_k^i \left\| \begin{array}{c} (-1,0),(0,0),(1,0) \\ \begin{bmatrix} 0 & 0 & 0 & 0 \\ 1 & 0 & 0 & 1 \\ 0 & 1 & 0 & 2 \\ 0 & 0 & 1 & 1 \end{bmatrix} \end{array} \right\| \right] \qquad (73)$$

which provides $N = 0.5$. On the other hand, rule 26, based on Table 4 and Eq. (21) can be readily identified as:

$$C_k^{i+1} = \tfrac{1}{3}\mathrm{I}\Delta_2^5 \left[ C_k^i \left\| \begin{array}{c} (-1,0),(0,0),(1,0) \\ \begin{bmatrix} 0 & 0 & 0 & 0 \\ 1 & 0 & 0 & 1 \\ 0 & 1 & 0 & 2 \\ 0 & 0 & 1 & 4 \end{bmatrix} \end{array} \right\| \right] \qquad (74)$$

which also gives $N = 0.5$. Even Neumann numbers are the same and both CA schemes are representative of pure diffusion, the behavior described is radically different as it can be seen on Figures 10 and 11.

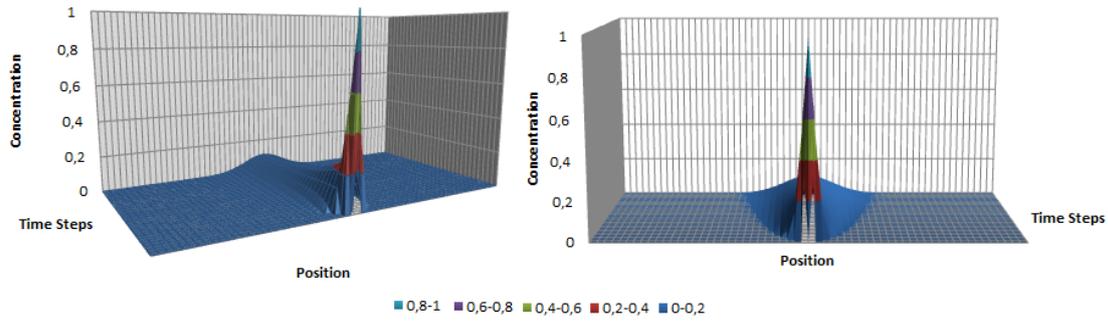

Figure 10. Rule 146 CA scheme

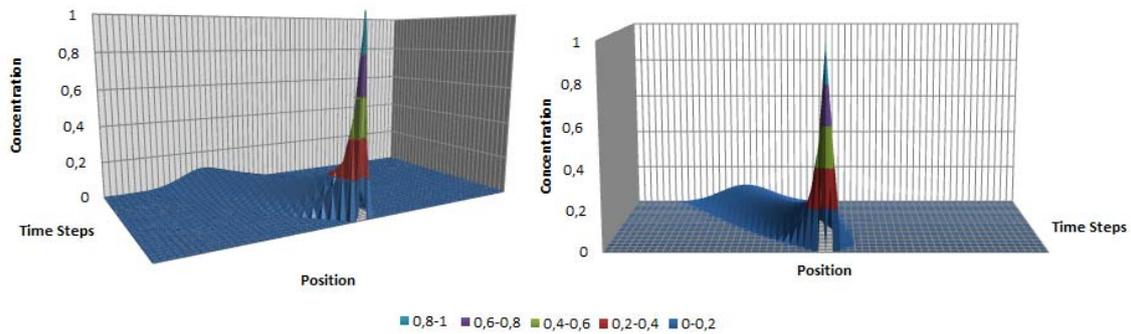

Figure 11. Rule 26 CA scheme

Figure 10 shows a symmetric diffusion while Figure 11 shows an asymmetric distribution whose peak is constantly moving. It is undeniable that anomalous diffusion takes place in the latter figure. This leads to a new understanding of how normal and anomalous diffusion are correlated. It can be said that normal diffusive processes are doubly symmetric while anomalous diffusion is only once symmetric. Correlating cellular automata and physical processes described by them to symmetry shows once more how valuable the latter is in science.

## Future Research Topics

The new way of describing cellular automata in terms of the iota-delta function has shown its value as a powerful tool of describing diffusive phenomena. On the other hand, other phenomena can be readily explored by means of the parallel between FDM schemes and CA.

Another interesting topic would be to investigate the relation between rules 30 and 110. Their iota-delta representations are closely related since, for the former, the generation rule is:

$$C_k^{i+1} = \tfrac{1}{3}\mathrm{I}\Delta_2^5 \left[ C_k^i \left\| \begin{array}{c} (-1,0),(0,0),(1,0) \\ \left\| \begin{array}{cccc} 0 & 0 & 0 & 0 \\ 1 & 0 & 0 & 1 \\ 0 & 1 & 0 & 4 \\ 0 & 0 & 1 & 4 \end{array} \right\| \end{array} \right. \right] \tag{75}$$

and for the latter:

$$C_k^{i+1} = \tfrac{1}{3}\mathrm{I}\Delta_2^5 \left[ C_k^i \left\| \begin{array}{c} (-1,0),(0,0),(1,0) \\ \left\| \begin{array}{cccc} 0 & 0 & 0 & 0 \\ 1 & 0 & 0 & 2 \\ 0 & 1 & 0 & 4 \\ 0 & 0 & 1 & 4 \end{array} \right\| \end{array} \right. \right] \tag{76}$$

Also, the usage of the iota-delta representation of cellular automata and specially rule 110, as in Eq. (76), to prove the universality in elementary cellular automata, as demonstrated by Cook (2004), is an interesting topic which would, for sure, demand some attention.

## Conclusion

Nature seems to be discrete; notwithstanding, the current scientific society is still dominated by the continuum idea. In order to empower the discrete notion, scientists have to develop methods to quali-quantitatively describe nature's behavior taking into account the discreteness in the latter. Cellular automata have shown to be an accurate description of some complex phenomena. In the present paper a general transformation which can be applied to the whole cellular net is developed. By means of such transformation, every binary, i. e., 0 – 255 cellular automata is described.

Besides, in order to provide a compact version of the transformation developed, a new function has been introduced: the iota-delta function. This new function is closely related to prime numbers and to the prime number theorem by means of the prime counting function, which once more proves the importance of this kind of numbers is science.

The iota-delta function is further generalized in order to describe every cellular automaton. Such generalization is the capital iota-delta function, which is related to set partition and the multinomial theorem. Finally, both the ordinary and the capital iota-delta functions provide an easy way to determine the value of any cell in the cellular automaton net, this way, some questions concerning random or chaotic cellular automata can be, for the first time, properly addressed. As an example, rule 30 has been analyzed and it has been shown that the new functions give straightforward ways to describe cellular automata.

By means of a correlation between the iota-delta function and the finite difference method, 0-255 cellular automata could be described as advective-dispersive processes. A new intuition has been brought up concerning normal and anomalous diffusion. It is worth noticing that such intuition could not have been deduced from PDEs and continuum theory since cellular automata modeling of nature was fundamental to its obtention.

The present paper intends to bring to discussion the iota-delta function and how the latter can be successfully applied to quantitatively describe cellular automata. Besides, the new intuition introduced seems to have interesting physical features which were not present in the previous definition of cellular automata. Also, some questions concerning cellular automata such as universality may be better analyzed by means of the iota-delta function.

## Acknowledgements

The authors would like to thank University of Brasilia, specially the Post Graduation Program in Geotechnics.